\documentclass[%
 reprint,
superscriptaddress,
 amsmath,amssymb,
 aps,prx,
floatfix,
]{revtex4-2}

\usepackage{braket}
\usepackage{mathrsfs}
\usepackage{graphicx}
\usepackage{dcolumn}
\usepackage{bm}
\usepackage{hyperref}
\hypersetup{pdfpagemode=FullScreen,
colorlinks=true, citecolor = blue, linkcolor=blue, filecolor=blue}
\usepackage[mathlines]{lineno}
\usepackage[T1]{fontenc}

\usepackage[colorinlistoftodos,bordercolor=red!50,backgroundcolor=red!10,linecolor=red!50,textsize=tiny]{todonotes}
\usepackage{changes}
\setaddedmarkup{\color{red} \uline{#1}}
\setdeletedmarkup{\color{red} \sout{#1}}
\sethighlightmarkup{{\color{blue} #1}}

\begin{document}

\preprint{APS/123-QED}

\title{Robust quantum control for the manipulation of solid-state spins}

\author{Yifan Zhang}
\altaffiliation[]{These authors contributed equally.}
\affiliation{%
Center for Quantum Technology Research and Key Laboratory of Advanced Optoelectronic Quantum Architecture and Measurements (MOE), School of Physics, Beijing Institute of Technology, Beijing 100081, China
}%
\affiliation{%
Beijing Academy of Quantum Information Sciences, Beijing 100193, China
}%

\author{Hao Wu}%
\altaffiliation[]{These authors contributed equally.}
\affiliation{%
Center for Quantum Technology Research and Key Laboratory of Advanced Optoelectronic Quantum Architecture and Measurements (MOE), School of Physics, Beijing Institute of Technology, Beijing 100081, China
}%
\affiliation{%
Beijing Academy of Quantum Information Sciences, Beijing 100193, China
}%

\author{Xiaodong Yang}
\altaffiliation[]{These authors contributed equally.}
\affiliation{Shenzhen Institute for Quantum Science and Engineering, Southern University of Science and Technology, Shenzhen, 518055, China}\affiliation{International Quantum Academy, Shenzhen 518048, China}
\affiliation{Guangdong Provincial Key Laboratory of Quantum Science and Engineering, Southern University of Science and Technology, Shenzhen, 518055, China}

\author{Ye-Xin Wang}
\affiliation{%
Spin-X Institute, School of Chemistry and Chemical Engineering, State Key Laboratory of Luminescent Materials and Devices, Guangdong-Hong Kong-Macao Joint Laboratory of Optoelectronic and Magnetic Functional Materials, South China University of Technology, Guangzhou 510641, China
}%

\author{Chang Liu}
\affiliation{%
Center for Quantum Technology Research and Key Laboratory of Advanced Optoelectronic Quantum Architecture and Measurements (MOE), School of Physics, Beijing Institute of Technology, Beijing 100081, China
}%
\affiliation{%
Beijing Academy of Quantum Information Sciences, Beijing 100193, China
}%

\author{Qing Zhao}
\affiliation{%
Center for Quantum Technology Research and Key Laboratory of Advanced Optoelectronic Quantum Architecture and Measurements (MOE), School of Physics, Beijing Institute of Technology, Beijing 100081, China
}%
\affiliation{%
Beijing Academy of Quantum Information Sciences, Beijing 100193, China
}%

\author{Jiyang Ma}
\email{mjy@bit.edu.cn}
\affiliation{%
Center for Quantum Technology Research and Key Laboratory of Advanced Optoelectronic Quantum Architecture and Measurements (MOE), School of Physics, Beijing Institute of Technology, Beijing 100081, China
}%
\affiliation{%
Beijing Academy of Quantum Information Sciences, Beijing 100193, China
}%

\author{Jun Li}
\email{lij3@sustech.edu.cn}
\affiliation{Shenzhen Institute for Quantum Science and Engineering, Southern University of Science and Technology, Shenzhen, 518055, China}\affiliation{International Quantum Academy, Shenzhen 518048, China}
\affiliation{Guangdong Provincial Key Laboratory of Quantum Science and Engineering, Southern University of Science and Technology, Shenzhen, 518055, China}

\author{Bo Zhang}
\email{bozhang_quantum@bit.edu.cn}
\affiliation{%
Center for Quantum Technology Research and Key Laboratory of Advanced Optoelectronic Quantum Architecture and Measurements (MOE), School of Physics, Beijing Institute of Technology, Beijing 100081, China
}%
\affiliation{%
Beijing Academy of Quantum Information Sciences, Beijing 100193, China
}%

\date{\today}

\begin{abstract}
Robust and high-fidelity control of electron spins in solids is the cornerstone for facilitating applications of solid-state spins in quantum information processing and quantum sensing. However, precise control of spin systems is always challenging due to the presence of a variety of noises originating from the environment and control fields. Herein, noise-resilient quantum gates, designed with robust optimal control (ROC) algorithms, 
are demonstrated experimentally with nitrogen-vacancy (NV) centers in diamond to realize tailored robustness against detunings and Rabi errors simultaneously. In the presence of both 10\% off-resonant detuning and deviation of a Rabi frequency, we achieve an average single-qubit gate fidelity of up to 99.97\%. Our experiments also show that, ROC-based multipulse quantum sensing sequences can suppress spurious responses resulting from finite widths and imperfections of microwave pulses, which provides an efficient strategy for enhancing the performance of existing multipulse quantum sensing sequences.
\end{abstract}

\maketitle

\section{Introduction}
Quantum optimal control (QOC) provides a powerful strategy to improve process performance in quantum technologies by designing efficient control fields against pulse errors in quantum operations. Fundamental quantum operations, such as state preparations, noise suppression and high-fidelity quantum gates \cite{design-high-fidelity-quantum-gate, design-NMR-pulse-by-gradient-algorithm}, play significant roles in magnetic-resonance spectroscopy and imaging \cite{QOC-for-MRI}, quantum information processing \cite{fuchs2011quantum,waldherr2014quantum,neumann2010quantum} and quantum sensing \cite{haberle2013high,shi2014sensing,shi2015single,aslam2017nanoscale,nano-MRI-via-three-level-control, nuclear-spin-gyroscope-based-on-NV}, which may benefit from recent advances of QOC \cite{rembold2020introduction}. 

To date, various methods have been developed to achieve high-fidelity operations such as composite pulses \cite{levitt1986composite}. However, for conventional composite pulses, such as broadband number 1 pulse (known as BB1 \cite{1994JMagR.109..221W}) and compensation for off-resonance errors with a pulse sequence (known as CORPSE \cite{PhysRevA.67.042308}), they are robust against a single type of pulse errors; for concatenated composite pulses, such as BB1inC \cite{rong2015experimental}, although it can simultaneously correct various types of existing errors but at the price of large pulse-widths.

\begin{figure} 
    \centering
    \includegraphics{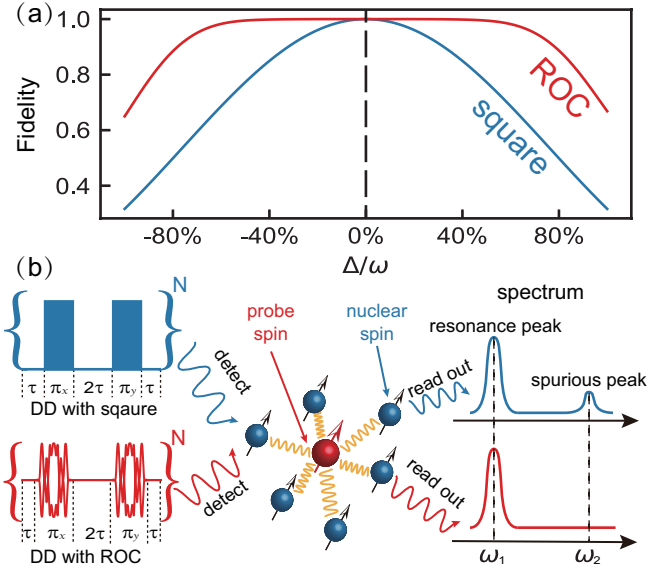}
    \caption{(a) Comparison of the robustness of square and ROC against pulse imperfections,  respectively. The blue (red) curve is simulated fidelities of square (ROC) $\pi$ pulses with detunings of MW frequency ranging from $-\Omega$ to $\Omega$, where $\Omega$ is the Rabi frequency. (b) Schematic of detecting and manipulating weakly coupled spins with dynamical decoupling (DD) sequences. Typical DD sequences (square) and robust DD sequences (ROC) are shown in the left part. The spectra derived from the two DD sequences are shown in the right part of (b). The real peak at $\omega_1$ results from surrounded nuclear spins and the spurious peak at $\omega_2$ results from the finite pulse length width and imperfections of square pulses. In contrast, the ROC pulses strongly suppress spurious peaks.}
    \label{fig:Conceptfig}
\end{figure}

In this letter, by exploiting robust optimal control (ROC) algorithms \cite{yang21, robust-optimal-control} in pulse-sequence designs, we experimentally demonstrate ROC as a simple, effective and hardware-friendly approach of quantum optimal control for realizing high-fidelity quantum gates and extensive robustness to multiple pulse errors. Consequently, the ROC approach suppresses pulse errors to the second order and results in the central robust region as flat and broad as possible, as shown in Fig.~1(a). Additionally, combining dynamical decoupling (DD) sequences with ROC, spurious peaks arising from finite pulse width and pulse imperfections can be suppressed, as shown in Fig.~\ref{fig:Conceptfig}(b).

\section{Quantification of system errors}
ROC-based high-fidelity single-qubit gates were demonstrated on a single electron spin of a negatively charged nitrogen-vacancy (NV) defect center in diamond. The ground state of NV center ($S=1$) is an electron spin triplet state with three sublevels $\ket{m_s=0}$ and $\ket{m_s=\pm1}$. The general Hamiltonian of the NV center with microwave (MW) control is 
\begin{equation}
\begin{split}
H_{\textrm{NV}} &= (\Delta+\delta_0)S_z\\
&+(\Omega+\delta_1)(\cos(\phi+\delta_\phi)S_x+\sin(\phi+\delta_\phi)S_y),
\end{split}
\end{equation}
where $\Delta$ is the detuning of MW control relative to the NV center's resonance frequency which we shall take as constant here, $\Omega$ and $\phi$ are the Rabi frequency and phase of MW pulse, respectively.. The corresponding errors resulting in gate errors are $\delta_0$, $\delta_1$ and $\delta_\phi$ respectively. $\delta_0$ represents static-field errors arising from the Overhauser field, magnetic fluctuations and unstable MW frequencies. $\delta_1$ consists of two parts, namely, (i) MW-amplitude errors mainly arising from static fluctuations of MW powers; (ii) errors of random noise caused by fluctuations of temperature and instability of radiation efficiency of coplanar waveguides. The phase error $\delta_\phi$ is attributed to imperfect MW generation.
\begin{figure}[htbp]
    \centering
    \includegraphics{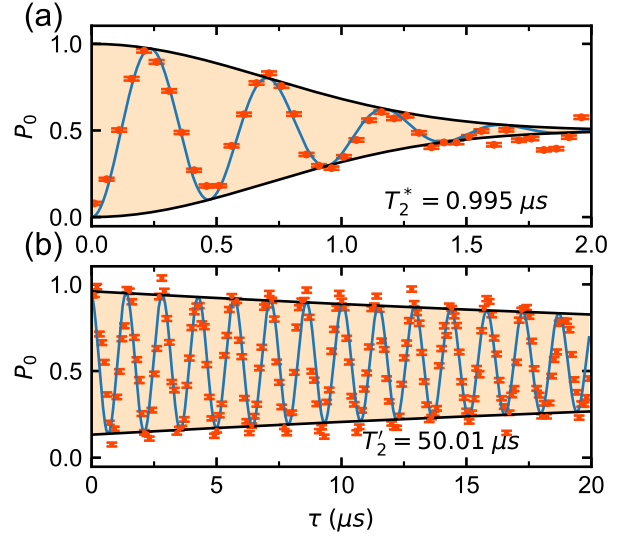}
    \caption{Characterization of the noises in experiments. (a) Results of the FID experiment, with corresponding pulse sequence in the form of $\frac{\pi}{2}-\tau-\frac{\pi}{2}$. Experimental data is shown as orange circles and the  blue solid line is the fit of the data with $f_0(\delta_0)$. The decay time of FID signal is $T_2^*=0.995$ $\mu s$. (b) Results of Rabi experiment. The increment of MW pulse length was set to be 100 ns. The experimental data is fit with function $f_1(\delta_1)$ (blue line). The decay time is $T_2'=50.01$ $\mu s$. Error bars on the data points are standard deviations from the mean.}
    \label{fig:quantify_error}
\end{figure}

After optimizing the microwave circuits (see Appendix~\ref{appendix:optimization MW pulses}), we quantified the errors resulting from the fluctuations of the experimental parameters. We assumed that the timescale of $\delta_0$ and $\delta_1$ was much longer than that of the single experiment, so the two errors were taken as quasi-static random contrasts. The distribution of $\delta_0$ was measured by free induction decay (FID) experiments with a pulse sequence in the form of $R_x(\frac{\pi}{2})-\tau-R_x(\frac{\pi}{2})$, in which $R_x(\frac{\pi}{2})$ was a rotation around $x$ axis by an angle $\frac{\pi}{2}$ under the rotating frame and $\tau$ was the free evolution time. As shown in Fig.~\ref{fig:quantify_error}(a), oscillatory FID signals were observed with the detuning of MW frequency $\Delta=2\pi\times2$ MHz. We assumed that $\delta_0$ satisfied a Gaussian distribution, so the probability distribution function is $f_0(\delta_0)=\frac{1}{\sigma\sqrt{2\pi}}e^{-\delta_0^2/2\sigma^2}$, where $\sigma$ stands for the standard deviation of the distribution.
The statistical probability that the final state $\ket{m_s=0}$ is preserved is
\begin{equation}
\begin{split}
    P & =\int_{-\infty}^{\infty}f(\delta_0)P_\mathrm{single}d\delta_0\\
    & =\frac{1}{2}-\frac{1}{2}e^{-\left(\frac{t}{T_2^*}\right)^2}\cos(2\pi\Delta t),
\end{split}
    \label{eq:quantify_errors}
\end{equation}
where $P_\mathrm{single}=\frac{1}{2}-\frac{1}{2}\cos(2\pi(\Delta+\delta_0)t)$ is the probability of state $\ket{m_s=0}$ in a single experiment, $T_2^*=1/(\sqrt{2}\pi\sigma)$ is the decay time of FID signals. The fitting results shown in Fig.~\ref{fig:quantify_error}(a) (the blue curve) illustrated that $\sigma=0.226(2)$ MHz and $T_2^*=0.995$ $\mu $s.
\begin{figure*}[htbp!]
\includegraphics{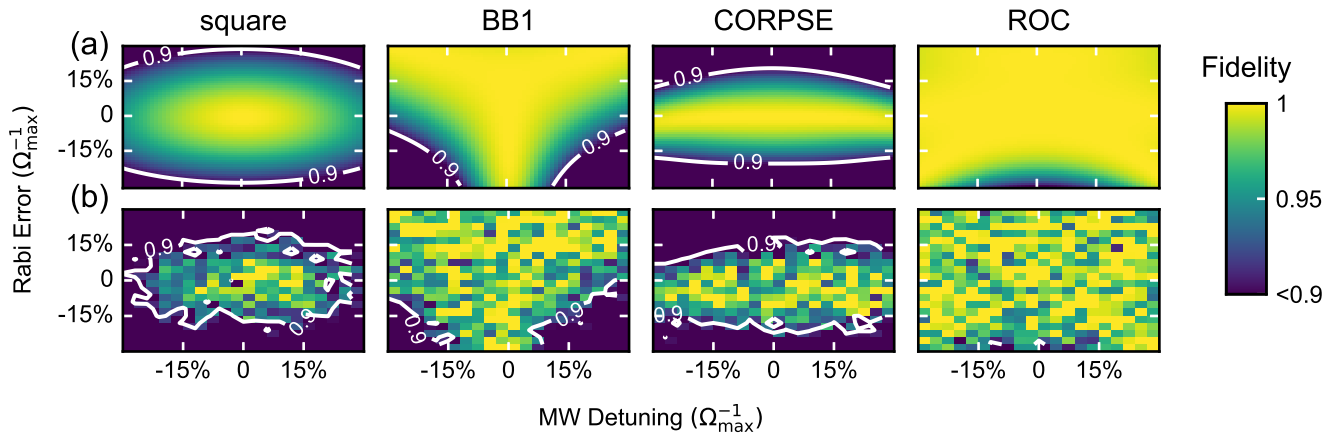}
\caption{State transfer fidelity of the four types of $\pi$-pulses compared. From left to right are square, CORPSE, BB1 and ROC respectively. Simulated ((a)) and measured ((b)) state transfer fidelity used the same maximum resonant Rabi frequency of $2\pi\times10$ MHz. White lines are contour lines at a fidelity of 0.9. The ranges of detunings and Rabi errors corresponded to $\pm 30 \%$ of the Rabi frequency. The averaged standard errors of the four types of pulses are 0.0318 (square), 0.0313 (CORPSE), 0.0321 (BB1) and 0.0309 (ROC), respectively, which lead to the fidelities of some random pixels lower than that of the surrounding pixels.}
\label{fig:2Dfig}
\end{figure*}
The error of Rabi frequency $\delta_1$ was quantified via Rabi experiments where $\delta_1$  satisfied a Lorentzian distribution of $f_1(\delta_1)=\gamma/(\pi(\delta_1^2+\gamma^2))$, and $\gamma$ stands for the half-width at half-maximum of the distribution and is inversely proportional to the decay time $T_2'$. The results of Rabi experiment are shown in Fig.~\ref{fig:quantify_error}(b). In order to quantify the MW field noise, we adjusted the length of MW pulses from 10 ns to 20 $\mu $s with an increment of 100 ns. The best fitting result (blue curve in Fig.~\ref{fig:quantify_error}(b)) was achieved with $T_2'=50.01$ $\mu $s.

\section{High-fidelity quantum gates}
We demonstrated high-fidelity single-qubit gates implementing four types of pulse sequences, which are square, BB1, CORPSE and ROC, respectively. Their pulse shapes are schematically depicted in Appendix~\ref{appendix:high-fidelity quantum gate}. Among different types of pulses, square pulses are widely used in quantum control protocols but are not sufficient robust to detunings and Rabi errors. CORPSE and BB1 are both composite pulses; the former can normally resist detunings and the latter is robust against Rabi errors. ROC is a robust shaped pulse optimized by ROC algorithms \cite{Haas_2019} which provides a general and effective way to enhance robustness of quantum controls against both MW-frequency detunings and Rabi errors. The procedure of generating a ROC pulse is described in Appendix~\ref{appendix:robust optimal control}. 

In order to experimentally verify that ROC is indeed robust with respect to detunings and Rabi errors, an NV electron spin initialized to $\ket{m_s=0}$ was irradiated with a $\pi$ pulse, whose carrier frequency was scanned over a range of $\pm2\pi\times3$ MHz across resonance and the amplitude was varied within a range of $\pm30\%$ of Rabi frequency. The $\pi$ pulse was generated based on the four different types of pulse shapes, i.e. square, BB1, CORPSE and ROC. The fidelity of $\pi$-pulse induced state transfer was defined by $F(\rho, \rho_0)=(\textrm{Tr}\sqrt{\rho^{1/2}\rho_0\rho^{1/2}})^2$, where $\rho$ and $\rho_0$ are the expected density matrix and the realistic density matrix, respectively. The experimental results shown in Fig.~\ref{fig:2Dfig}(b) agrees well with the theoretical fidelity landscapes (see Fig.~\ref{fig:2Dfig}(a)). Among the four types of pulses, it is evident that ROC-based $\pi$ pulse outperforms others in terms of the fidelities over a wide range of detunings and control amplitudes (both of them were more than $\pm30\%$ of Rabi frequency).
\begin{figure}[htbp!]
\centering
\includegraphics[]{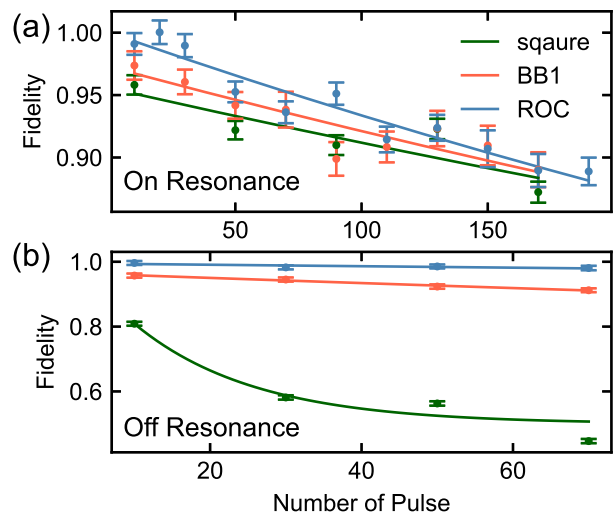}
\caption{(a) Results of resonant randomized benchmarking for a single-qubit. Points in the figure represent average fidelities of the final states obtained from each of the individual sequences of gates. The error bars on the data points are standard deviations from the mean. Solid lines are fit to experimental data using Eq.~\ref{eq:rb eq}. The averaged $\pi$ pulse fidelities of square, BB1 and ROC pulses are 0.9994(1), 0.99942(9) and 0.99928(7) respectively. (b) Results of off-resonant randomized benchmarking. Both MW detunings and Rabi errors are $10\%$ of $\Omega$, with $\Omega=2\pi\times10$ MHz. The averaged gates fidelities of square, BB1 and ROC pulses are 0.96(1), 0.99909(7) and 0.9997(1), respectively.}
\label{fig:RB}
\end{figure}

\section{Randomized Benchmarking}
To quantify the fidelity of a quantum gate, one method is the quantum process tomography \cite{quantum-process-tomography}. However, standard process tomography is limited by errors in the process of state preparation, measurement and single-qubit gates \cite{PhysRevA.77.012307}. In addition, due to the fact that error probabilities for quantum gates are of the order of 0.0001 or lower,  experimentally detecting such low errors is challenging. Here we use the randomized benchmarking (RB) method \cite{PhysRevA.77.012307} to determine average gate fidelities. In an RB experiment, gate fidelities are evaluated by measuring the fidelity of the final state after random sequences are applied. Errors of state preparations and measurements are separated thus the gate fidelity is determined precisely. The qubit is initialized to $\ket{0}$ state, then a predetermined sequence of randomized computational gates is applied. Each computational gate consists of a Pauli gate and a Clifford gate which are in the form of $e^{\pm i\sigma_a \pi/2}$ and $e^{\pm i \sigma_b \pi/4}$ respectively, where $a$ is chosen randomly in $\{0,x,y,z\}$ with an identity gate $\sigma_0$, and $b$ is chosen randomly in $\{x,y\}$ (see Appendix~\ref{appendix:rb}). As the number of computational gates increases, the accumulation of gate errors reduces the measured fidelities of final states. The average fidelity of final states is given by \cite{PhysRevA.77.012307}:
\begin{equation}
    \bar{F}=\frac{1}{2}+\frac{1}{2}(1-d_{if})(2F_a-1)^l
    \label{eq:rb eq}
\end{equation}
where $d_{if}$ is errors of readout and state preparations, $F_a$ is the average fidelity of the NOT gate and $l$ is the number of randomized computational gates.

We initialized a single NV spin in $\ket{m_s=0}$ and demonstrated RB experiments using three types of pulses (square, BB1 and ROC) under resonant (Fig.~\ref{fig:RB}(a)) and off-resonant conditions (Fig.~\ref{fig:RB}(b)) to quantify gate fidelities precisely. The Rabi frequency was set to be $\Omega=2\pi\times10$ MHz. Under the resonant condition, the average gate fidelities were 0.9994(1) for square, 0.99942(9) for BB1 and 0.99928(7) for ROC respectively, showing no advantages of ROC. While under the off-resonant condition in which detunings and Rabi errors were both set to be $10\%$ of $\Omega$, the highest fidelity of 0.9997(1) was achieved when ROC pulses were applied.

\section{Detection of nuclear spins}
The ROC algorithm is not only effective   in designing robust quantum gates, but can also be applied to improve    dynamical decoupling sequences in  the task  of detecting  weakly  coupled nuclear spins. Nuclear spins exist naturally in abundance in diamonds functioning as additional quantum resources \cite{RN10, PhysRevX.9.031045}. A prerequisite to exploit nuclear spins' quantum properties is to detect and characterize the nuclear spins, which can be realized by utilizing the electron spins \cite{PhysRevLett.109.137602}. The Hamiltonian of the NV spin system under the rotating frame is given by \cite{PhysRevLett.109.137602}:
\begin{equation}
\begin{split}
        H=\overbrace{\Omega(x(t)S_x+y(t)S_y)}^{H_\mathrm{C}}+\overbrace{a_{\parallel}S_z I_z+a_\perp S_z I_x+\omega_\mathrm{l} I_z}^{H_\mathrm{free}}
\end{split}
    \label{eq:hamiltonian}
\end{equation}
where $H_\mathrm{C}$ and $H_\mathrm{free}$ are the Hamiltonians of control fields and free evolution respectively. $x(t)$ and $y(t)$ are amplitude modulations of MW pulses, $a_{\parallel}(a_\perp)$ represents the parallel (transverse) component of hyperfine coupling strength $\omega_\mathrm{h}$, and $\omega_\mathrm{l}$ is the Larmor frequency of nuclear spins, $S_x$, $S_y$ and $S_z$ are electron spin operators, $I_x$ and $I_z$ are nuclear spin operators. The Hamiltonian of free evolution can be reformed as:
\begin{equation}
    H_\mathrm{free}=\ket{0}\bra{0}\omega_\mathrm{l} I_z+\ket{1}\bra{1}[(\omega_\mathrm{l}+a_{\parallel})I_z+a_\perp I_x],
\end{equation}
where $\ket{0}(\ket{1})$ is the electron spin state of $\ket{m_s=0}(\ket{m_s=-1})$. It is clear that nuclear spin evolves conditionally according to the electron spin state. Since we performed the experiment at $B_0=510$ Gauss, under the approximation that $\omega_l\gg\omega_h$, the effective Larmor frequency of nuclear spins is $\omega_0=\omega_l+a_{\parallel}/2$. Thus the effects of the surrounding nuclear spin bath can be considered as an ac signal whose frequency is $\omega_0$ and amplitude is proportional to $a_\perp$ acting on the electron spin \cite{PhysRevLett.109.137602}.

\begin{figure}[b]
    \centering
    \includegraphics{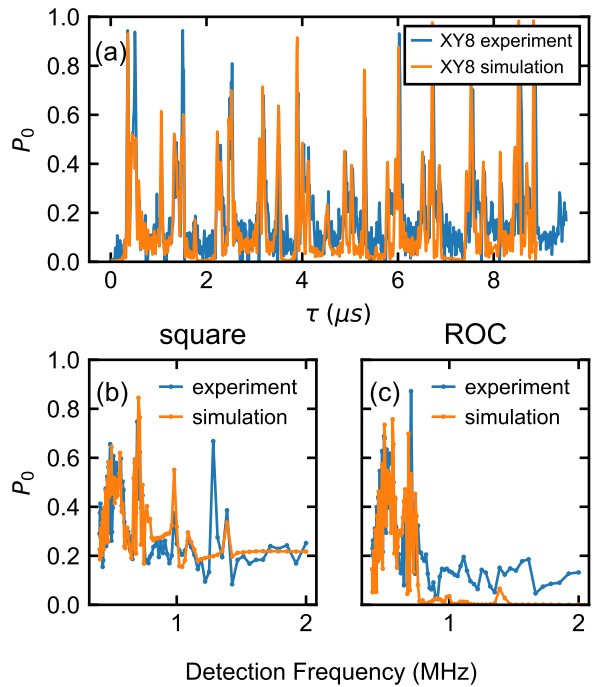}
    \caption{(a) Experiment (blue line) and simulation (orange line) of detecting nuclear spins with XY8-2 sequence. The Rabi frequency was $2\pi\times10$ MHz and magnetic field was $B_0=510$ Gauss. The simulation was based on five $^{13}\mathrm{C}$ nuclear spins coupled with a NV center and their coupling strengths were shown in Tab.~\ref{tab:TAB1}. (b) and (c) Experiments and simulations of detecting nuclear spins with XY4-40 sequences implemented with square ((b)) and ROC ((c)) under an off-resonant condition. Both detunings and Rabi errors were set to $8\%\times\Omega$. The detection frequency was $f=1/(4\tau)$. It can be clearly seen that there is a spurious peak at 1.39 MHz shown in (b), and the corresponding resonance peak was at 0.69 MHz, which was induced by spin 1 in Table.~\ref{tab:TAB1}.}
    \label{fig:DD}
\end{figure}
\begin{table}[b]
    \centering
    \begin{ruledtabular}
        \begin{tabular}{ccc}
         Spin & $\omega_h/2\pi$ (kHz) & $\theta$ (degrees)\\\hline
         1&360(9)&56(2)\\
         2&46(1)&178.3(5)\\
         3&152(5)&134(2)\\
         4&107(1)&122(1)\\
         5&67(4)&26(1)
    \end{tabular}
    \end{ruledtabular}
    \caption{Hyperfine coupling strengths $\omega_h$ and polar angles $\theta$ for four nuclear spins identified in Fig.~\ref{fig:DD}. For each nuclear spin these values were obtained by individually fitting a single well isolated resonance.}
    \label{tab:TAB1}
\end{table}

By applying DD sequences (with a basic unit in the form of $\tau-\pi-2\tau-\pi-\tau$) on the electron spin, a resonant peak indicating the coupling between the NV electron spin and the nuclear spin should occur when the interpulse delay $2\tau=k\pi/\omega_0$ ($k$ is odd). However, in realistic experimental implementation, each $\pi$ pulse has a finite width and imperfections, thus when interpulse delay satisfies $2\tau=k\pi/(2\omega_0)$, phase accumulation occurs during $\pi$ pulses \cite{PhysRevX.5.021009, Pasini_2011}. The phase accumulation at the second harmonic grows as the number of $\pi$ pulses increases and introduces spurious peaks in the detected spectrum, which leads to difficulties in identifying resonance peaks. To avoid false identifications, substantial robust DD sequences have been proposed \cite{PhysRevLett.118.133202, PhysRevLett.106.240501, PhysRevLett.122.200403, PhysRevA.92.042304,PhysRevLett.117.130502}. When the pulse imperfections are relatively large, our previous theoretical work \cite{yang21} indicates that the performance of reported robust DD sequences can be further improved by combining with ROC pulses, resulting in a rather distinguishable spectrum.

 Here, we first prepared a single NV electron spin in a superposition $\ket{x}=\frac{1}{\sqrt{2}}(\ket{0}-i\ket{1})$ by applying a rotational operation about the $x$ axis by an angle $\pi/2$. After applying the XY8 sequence \cite{GULLION1990479} twice (with a basic unit in the form of $\tau-\pi_x-2\tau-\pi_y-2\tau-\pi_x-2\tau-\pi_y-2\tau-\pi_y-2\tau-\pi_x-2\tau-\pi_y-2\tau-\pi_x-\tau$), another rotation about the $x$ axis by an angle $\pi/2$ prepared the state to $\ket{1}$. The resulted population in $\ket{1}$ as a function of $\tau$ was shown in Fig.~\ref{fig:DD}(a). According to the positions and depths of the harmonics, we found five $^{13}\mathrm{C}$ coupled with the electron spin. The coupling strengths and their polar angles were shown in Table.~\ref{tab:TAB1}. In order to observe the spurious harmonics, we chose the XY4 sequence \cite{GULLION1990479} (with a basic unit in the form of $\tau-\pi_x-2\tau-\pi_y-2\tau-\pi_x-2\tau-\pi_y-\tau$) implemented with square and ROC under an off-resonant condition. Both MW detuning and Rabi error were set to be $8\%\times\Omega$. By repeating the sequence for 40 times, Fig.~\ref{fig:DD}(b) shows a resonant peak at 0.695 MHz and a spurious peak at 1.39 MHz, which was induced by a strongly coupled nuclear spin (spin 1. in Tab) with its $a_{\parallel}=305$ kHz and $a_\perp=136$ kHz (Spin 4. in Tab.~\ref{tab:TAB1}). In comparison, by applying ROC algorithms in DD sequences, the background noise and spurious peaks caused by detunings and Rabi errors can be suppressed significantly, giving rise to a much more clear spectrum (Fig.~\ref{fig:DD}(c)) revealing the electron-nuclear couplings.

\section{summary}
Noise-resilient quantum gates devised by the robust optimal control algorithm were demonstrated experimentally with considerable improvement of gate fidelities. Unlike hard pulses, shaped pulses with smooth constraints are more friendly to hardware implementation. By generating quantum gates with ROC pulses, the effects of inhomogeneous MW detunings and fluctuations of MW powers were suppressed. In addition, combining DD sequence with ROC pulses suppressed all spurious peaks to a satisfactory extent. Furthermore, this presented approach is fully compatible with other quantum systems, such as superconducting qubits, trapped ions and Rydberg atoms. It would also be an essential technique to be added to the toolbox for improving the sensitivity of solid-state ensemble spin sensors, such as NV centers in diamond, V$_{Si}$ in SiC and organic spin sensors \cite{wu2022enhanced}, of which the sensitivities are limited by the inhomogeneous broadening \cite{nobauer2015smooth} or the dynamic biological environment \cite{konzelmann2018robust}.

\begin{acknowledgments}
We thank Tianyu Xie for stimulating discussions. This study was supported by NSF of China (Grant No. 12004037, No. 91859121), Beijing Institute of Technology Research Fund Program for Young Scholars and the China Postdoctoral Science Foundation (Grant No. YJ20210035, No. 2021M700439).  X.Y. and J.L. are supported by   the National Natural Science Foundation of China (Grants No. 1212200199, No. 11975117, and No. 92065111), Guangdong Basic and Applied Basic Research Foundation (Grant No. 2021B1515020070),  Guangdong Provincial Key Laboratory (Grant No. 2019B121203002), and Shenzhen Science and Technology Program (Grants No. RCYX20200714114522109  and   No. KQTD20200820113010023).
\end{acknowledgments}

\appendix
\section{Optimization of MW Pulses}
\label{appendix:optimization MW pulses}
\begin{figure*}[htbp]
    \centering
    \includegraphics{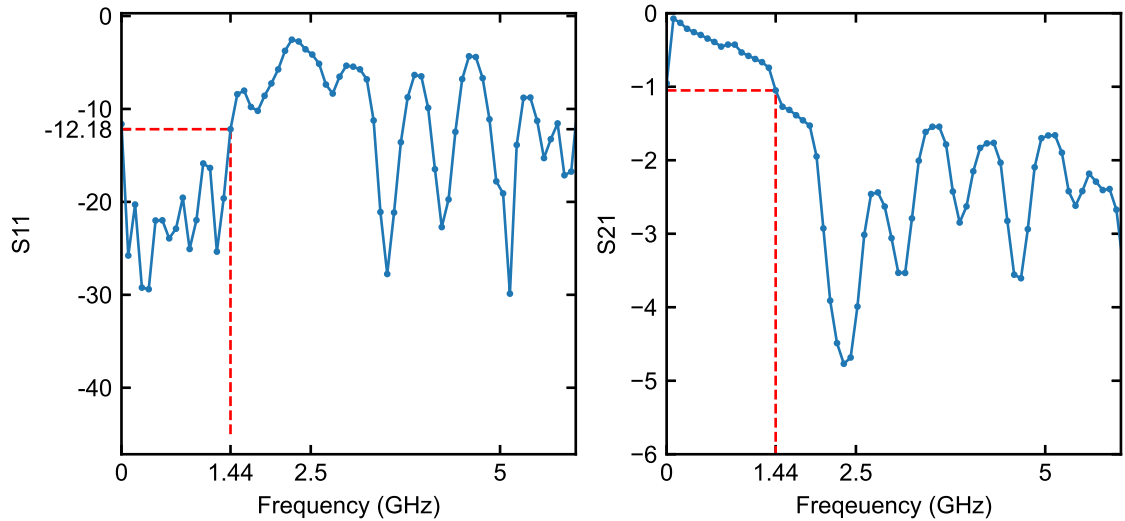}
    \caption{Scattering parameters of the CPW. The S11 and S21 parameters at 1.44 GHz are -12.18 dB and -1.05 dB respectively.}
    \label{fig:Sparameter}
\end{figure*}
\begin{figure*}[htbp]
    \centering
    \includegraphics{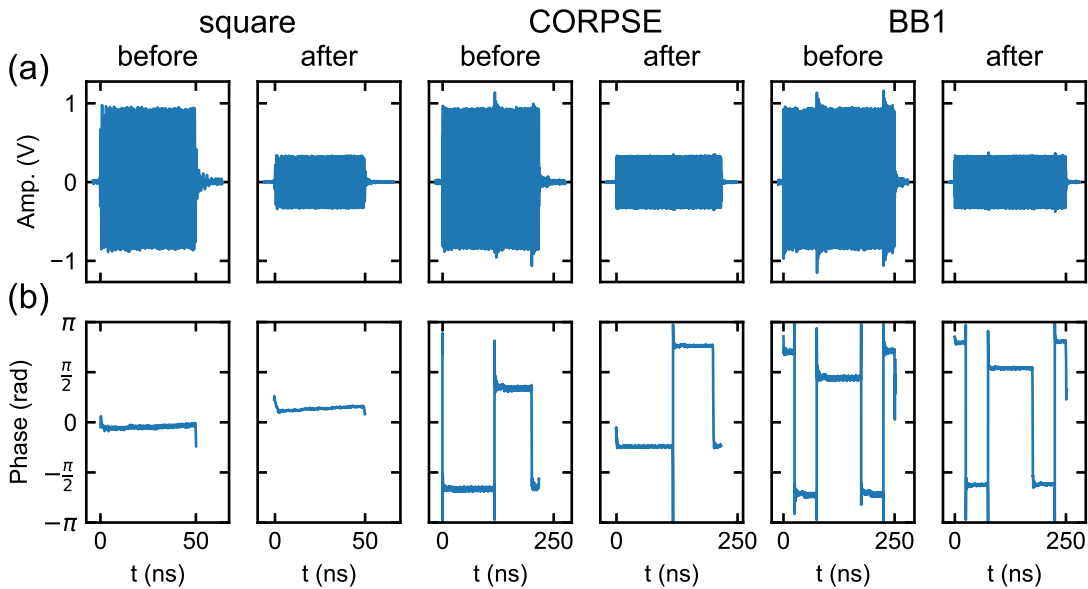}
    \caption{Optimization of microwave pulses. (a) Waveforms of square, CORPSE and BB1 before and after the optimization. (b) Results of phase demodulation before and after the optimization.}
    \label{fig:wave_analysis}
\end{figure*}
\begin{figure*}[htbp]
    \centering
    \includegraphics{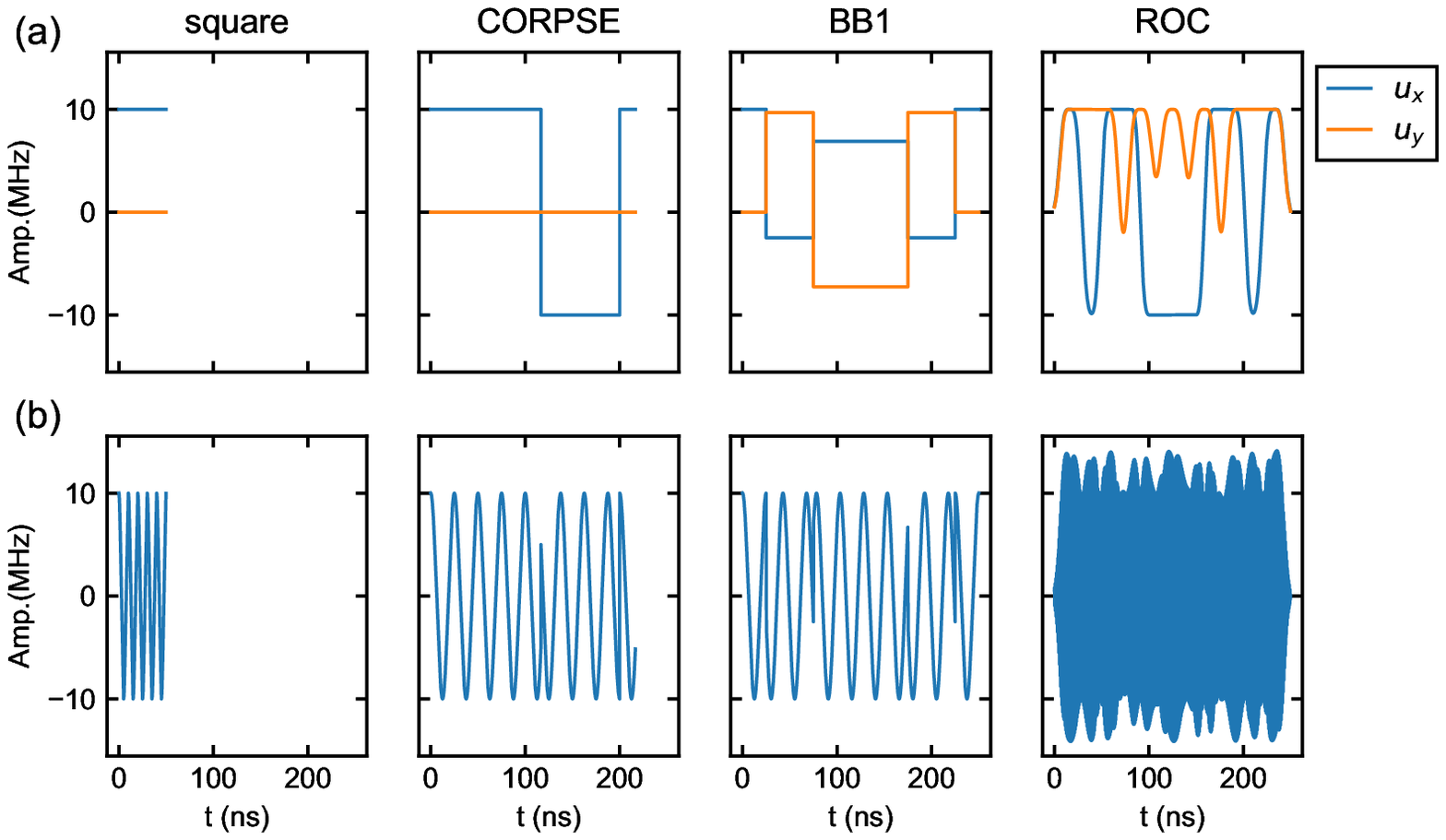}
    \caption{(a)Amplitudes of in-phase components (blue lines) and quadrature components (yellow lines) of $(\pi)_0$ gates. (b) Waveform schematics of square, CORPSE, BB1 and ROC.}
    \label{fig:waveform_schematic}
\end{figure*}
In our experiments, manipulations of qubits were achieved by microwave pulses, which were generated by an arbitrary wave generator (AWG70001A, Tektronix), amplified with a power amplifier (ZHL-15W-422-S+, Mini Circuits) and applied through a coplanar waveguide (CPW). In order to minimize the effects induced by both the Overhauser field (caused by interactions with nuclear spin bath) and the distortion of microwave amplitudes (caused by limited bandwidth), we designed and fabricated an ultra-broadband CPW structure with a bandwidth up to 6 GHz \cite{jia2018ultra}. The scattering parameters of the CPW shown in Fig.~\ref{fig:Sparameter} were measured by a microwave analyzer (N9917A, KeySight). 

We found that leakages and reflections between microwave components resulting in extra distortions of microwave pulses, which decreased the fidelity of quantum gates. Figure.~\ref{fig:wave_analysis}(a) shew the waveforms of square, CORPSE and BB1. It was apparent that there were distortions at the beginning and the end of pulses. For composite pulses, there were additional distortions between each pulses. Additionally,  noise and distortions of phase were found in these pulses (Fig.~\ref{fig:wave_analysis}(b)). In order to suppress the distortions and noises, we inserted a 6-dB attenuator at the output port of the microwave source. And at the output port of the power amplifier, we inserted a 6-dB attenuator and an isolator with its frequency bandwidth ranging from 1.42 to 1.54 GHz. Comparisons of the waveforms and phases obtained before and after the circuitry optimization were shown in Fig.~\ref{fig:wave_analysis}. The amplitude distortions and phase noises were successfully reduced.

\section{High-Fidelity Quantum Gate}
\label{appendix:high-fidelity quantum gate}
We denoted a single-qubit gate as $(\theta)_\phi$ corresponding to a rotation in the rotating frame of angle $\theta$ around the axis in the equatorial plane with azimuth $\phi$. Waveforms of four types of $(\pi)_0$ gate are shown in Fig.~\ref{fig:waveform_schematic}(b). The CORPSE \cite{03PRA} was depicted as $(\theta_1)_0-(\theta_2)_\pi-(\theta_3)_0$, with rotation angles $\theta_1=2\pi+\frac{\theta}{2}-\arcsin(\frac{\sin(\theta/2)}{2})$, $\theta_2=2\pi-2\arcsin(\frac{sin(\theta/2)}{2})$ and $\theta_3=\frac{\theta}{2}-\arcsin(\frac{\sin(\theta/2)}{2})$. The pulse sequence of BB1 \cite{WIMPERIS1994221} was $(\frac{\theta}{2})_0-(\pi)_\phi-(2\pi)_{3\phi}-(\pi)_\phi-(\frac{\theta}{2})_0$, where $\phi=\arccos(-\theta/4\pi)$. ROC was generated by robust optimal control methods \cite{yang21}. Its amplitude and phase varied with time. The in-phase components $u_x$ and quadrature components $u_y$ of the four types of pulses are shown in Fig.~\ref{fig:waveform_schematic}(a). A $\pi$ pulse with an arbitrary phase angle $\phi$ can be described with $(u_x\cos\phi-u_y\sin\phi)\hat{S_x}+(u_x\sin\phi+u_y\cos\phi)\hat{S_y}$.

\section{Normalization of Experimental Data}
\label{appendix:normalization data}
\begin{figure}[htbp]
    \centering
    \includegraphics{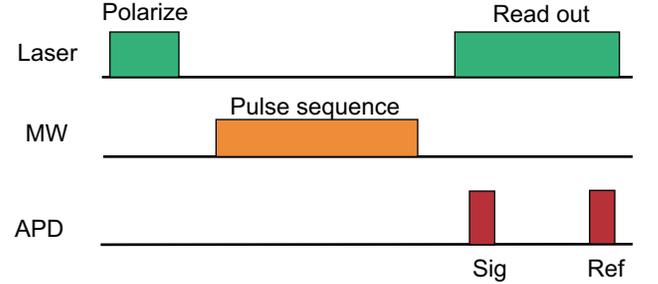}
    \caption{General experiment schematic}
    \label{fig:general_exp_shcematic}
\end{figure}
The general experiment schematic is shown in Fig.~\ref{fig:general_exp_shcematic}. A 532-nm green laser (MGL-III-532, EFORCE LASERS) was used to polarize the NV center to $\ket{m_s=0}$. Then the pulses generated with the AWG70001A were applied on the qubit. After the microwave pulses, another 532-nm laser pulse was used for readout, and the fluorescence was recorded by the CIQTEK quantum diamond single spin spectrometer. The first photon-counting event recorded as $I_\mathrm{sig}$ is labelled as 'Sig' in Fig.~\ref{fig:general_exp_shcematic}. The second photon-counting event ('Ref' in Fig.~\ref{fig:general_exp_shcematic}) recorded as $I_\mathrm{ref}$ is the reference. The experimental data was calculated by $C=\frac{\bar{I}_\mathrm{sig}-\bar{I}_\mathrm{ref}}{\bar{I}_\mathrm{ref}}$, where $\bar{I}_\mathrm{sig}(\bar{I}_\mathrm{ref})$ stands for the averaged value of $I_\mathrm{sig}(I_\mathrm{ref})$ collected in $8\times10^5$ datasets. The error for each experimental data can be derived according to
\begin{equation}
    \Delta C=\frac{\Delta I_\mathrm{sig}\cdot\bar{I}_\mathrm{ref}+\Delta I_\mathrm{ref}\cdot\bar{I}_\mathrm{sig}}{\bar{I}_\mathrm{sig}^2}
\end{equation}
where $\Delta \bar{I}_\mathrm{sig}$ and $\Delta \bar{I}_\mathrm{ref}$ denote standard deviations of $\bar{I}_\mathrm{sig}$ and $\bar{I}_\mathrm{ref}$ respectively. In general, a quantum state tomography \cite{quantum-process-tomography, quantum-state-tomo2} is normally used to get fidelity of final state. However, in our experiments, final states were either $\ket{m_s=0}$ or $\ket{m_s=-1}$, so we only needed to figure out the diagonal elements of the density matrix. Normalization of the experimental data was carried out by performing Rabi experiments. By fitting the data obtained from Rabi experiments with a sinusoidal function, the probability of achieving state $\ket{m_s=0}$ was given by:
\begin{equation}
    P_{\ket{m_s=0}}=\frac{C-C_\mathrm{min}}{C_\mathrm{max}-C_\mathrm{min}}
\end{equation}
where $C_\mathrm{min}$ and $C_\mathrm{max}$ were the maximum and minimum values of the fitted curve.
\begin{figure}[htbp!]
    \centering
    \includegraphics{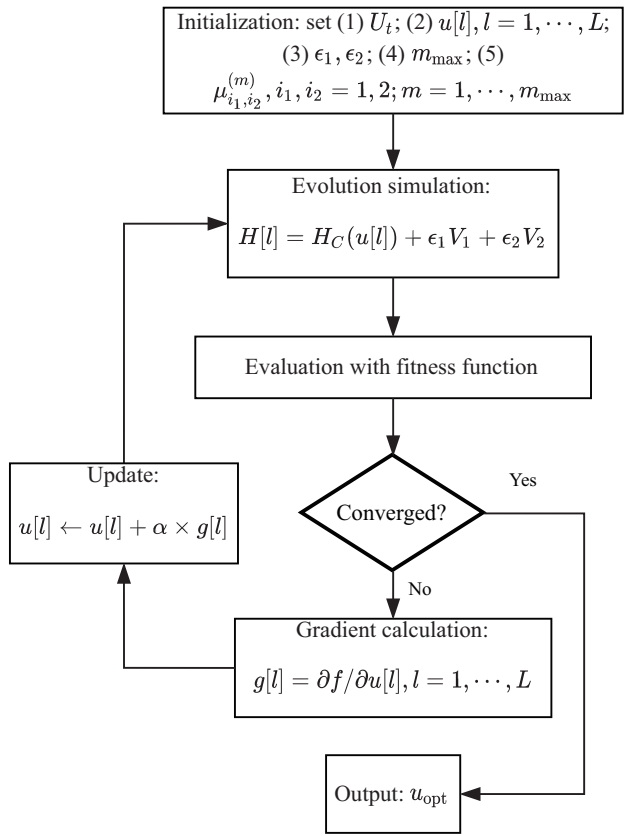}
    \caption{Flow chart of generating pulses with ROC method.}
    \label{fig:generateROC}
\end{figure}
\section{Robust Optimal Control}
\label{appendix:robust optimal control}
The robust optimal control method that we adopted in our work is based on Ref. \cite{yang21}. Here we plot a flow chart (see Fig.~\ref{fig:generateROC}) to show the procedure of generating a ROC pulse. The procedure starts from initializing some parameters including (1) the operator $U_t$ to be optimized (here we take $U_t=(\pi)_0$); (2) the pulse length $L$; (3) two weights of noise, $\epsilon_1$ for detunings and $\epsilon_2$ for Rabi errors; (4) the maximum order of directional derivatives $m_\mathrm{max}$; (5) a set of weights specifying the importance of the associated objective function $\mu_{i_1,i_2}^{(m)}$. Then according to the results of evolution simulation, the fitness function is calculated by:
\begin{equation}
\begin{split}
    \Phi(u) & =\lvert\mathrm{Tr}(U_C(\tau_u)R_0^\dagger(\theta))\rvert^2-\\
    & \sum_{m=1}^{m_\mathrm{max}}\sum_{i_1,i_2=1}^2\mu_{i_1,i_2}^{(m)}\lVert\mathscr{D}_{U_C}^{(m)}(V_{i_1}, V_{i_2})\rVert^2.
\end{split}
\end{equation}
If $\Phi$ is sufficiently high, the shaped pulse $u$ is output, otherwise, the gradient of the fitness function with respect to the pulse parameters $g[l]=\partial\Phi/\partial u[l]$ is computed. By determining an appropriate step length $\alpha$ along the search direction $g[l]$, we update the pulse parameters $u[l]\leftarrow u[l]+\alpha\times g[l]$ and go back to the simulation step.

\section{Randomized Benchmarking}
\label{appendix:rb}
The randomized benchmarking (RB) method \cite{PhysRevA.77.012307} consists of a large number of experiments. Each experiment includes the preparation of an initial state $\ket{m_s=0}$, application of an alternating sequence of "Pauli gate" and "Clifford gate", and measuring final states. The Pauli gates consist of unitary operators which  are in the form of $e^{\pm i \sigma_b \pi/2}$, where $b=0, x,y$ and $z$ and $\sigma_0$ is defined as an identity operator. The Clifford gates are $\pi/2$ pulses represented by $e^{\pm i \sigma_u \pi/4}$, with $u=x,y$. The sign and $u$ are chosen uniformly at random except for the last $\pi/2$ pulse, which is chosen to ensure that the final state is either $\ket{m_s=0}$ or $\ket{m_s=-1}$. The length $l$ of a randomized pulse sequence is its number of the $\pi/2$ pulses. By performing the experiment $N$ times for each length $l=1,\cdots,L$, the averaged state fidelity $\bar{F}$ is estimated. The average gate fidelity $F_a$ can be derived from the relationship between $l$ and $\bar{F}$, which is given by \cite{PhysRevA.77.012307}
\begin{equation}
    \bar{F}=\frac{1}{2}+\frac{1}{2}(1-d_{if})(2F_a-1)^l
\end{equation}
where $d_{if}$ combines the errors of the preparation and measurement, and $F_a$ represents the average fidelity of the NOT gate.

Measurement statistics for $N_G N_l N_P N_e$ experiments is obtained with the RB method, where $N_G$ is the number of different computational gate sequences, $N_l$ is the number of lengths to which the sequences are truncated, $N_P$ is the number of Pauli randomizations for each gate sequence, and $N_e$ is the number of experiments for each specific sequence. The steps of generating randomized sequences are given by \cite{PhysRevA.77.012307}:
\begin{enumerate}
    \item Pick a set of lengths $l_1<l_2<\cdots<l_{N_l}$ and do the following for each $j=1,2,\cdots, N_G$:
    \begin{enumerate}
        \item Choose a random sequence $\mathcal{G}=\{G_1,\cdots\}$ of $l_{N_l}-1$ computational gates.
        \item For each $k=1,\cdots,N_l$, do the following:
        \begin{enumerate}
            \item Figure out the final state $\rho_f$ by applying ideal sequence $G_{l_k}\cdots G_1 $ to $\ket{0}$.
            \item Randomly pick a final gate $R$ among the two $\pm x, \pm y, \pm z$ axis $\pi/2$ pulses applied to $\rho_f$ to ensure the final state is an eigenstate of $\sigma_z$.
            \item Do the following for each $m=1,\cdots N_P$:
            \begin{enumerate}
                \item Choose a random sequence $\mathcal{P}=\{P1,\cdots\}$ of $l_k+2$ Pauli pulses.
                \item Generate the sequence as $P_{l_k+2} R P_{l_k+1} G_{l_k}\cdots G_1 P_1$. Record the idea final state when applying this sequence to $\ket{0}$. Experimentally demonstrate the sequence on $\ket{0}$, repeating $N_e$ times.
                \item From the experimental data and the expected final state, obtain an estimate $F_{j,l_k,m}$ of the state fidelities.
            \end{enumerate}
        \end{enumerate}
    \end{enumerate}
\end{enumerate}

\bibliography{Ref}

\end{document}